\documentclass[%
 reprint,
 superscriptaddress,
 amsmath,amssymb,
 aps,
 prl,
]{revtex4-1}

\usepackage[pdftex]{graphicx}% Include figure files
\usepackage{dcolumn}% Align table columns on decimal point
\usepackage{bm}% bold math
\usepackage[colorlinks,linkcolor=blue,citecolor=blue]{hyperref}
\begin{document}

%\preprint{APS/123-QED}

\title{Observation of the Magnon Polarization}

\author{Y. Nambu}
 \email{nambu@tohoku.ac.jp}
 \affiliation{Institute for Materials Research, Tohoku University, Sendai 980-8577, Japan}
\author{J. Barker}
 \affiliation{Institute for Materials Research, Tohoku University, Sendai 980-8577, Japan}
 \affiliation{School of Physics and Astronomy, University of Leeds, Leeds LS2 9JT, UK}
\author{Y. Okino}
 \affiliation{Institute for Materials Research, Tohoku University, Sendai 980-8577, Japan}
\author{T. Kikkawa}
 \affiliation{Institute for Materials Research, Tohoku University, Sendai 980-8577, Japan}
 \affiliation{WPI-AIMR, Tohoku University, Sendai 980-8577, Japan}
\author{Y. Shiomi}
 \affiliation{Institute for Materials Research, Tohoku University, Sendai 980-8577, Japan}
\author{M. Enderle}
 \affiliation{Institut Laue-Langevin (ILL), 38042 Grenoble, France}
\author{T. Weber}
 \affiliation{Institut Laue-Langevin (ILL), 38042 Grenoble, France}
\author{B. Winn}
 \affiliation{Oak Ridge National Lab (ORNL), Oak Ridge, TN 37831, USA}
\author{M. Graves-Brook}
 \affiliation{Oak Ridge National Lab (ORNL), Oak Ridge, TN 37831, USA}
\author{J.M. Tranquada}
 \affiliation{Brookhaven National Lab (BNL), Upton, NY 11973-5000, USA}
\author{T. Ziman}
 \affiliation{Institut Laue-Langevin (ILL), 38042 Grenoble, France}
 \affiliation{Universit\'e Grenoble Alpes, CNRS, LPMMC, 38000 Grenoble, France}
\author{M. Fujita}
 \affiliation{Institute for Materials Research, Tohoku University, Sendai 980-8577, Japan}
\author{G.E.W. Bauer}
 \affiliation{Institute for Materials Research, Tohoku University, Sendai 980-8577, Japan}
 \affiliation{WPI-AIMR, Tohoku University, Sendai 980-8577, Japan}
 \affiliation{Zernike Institute for Advanced Materials, University of Groningen, 9747 AG Groningen, The Netherlands}
\author{E. Saitoh}
 \affiliation{Institute for Materials Research, Tohoku University, Sendai 980-8577, Japan}
 \affiliation{WPI-AIMR, Tohoku University, Sendai 980-8577, Japan}
 \affiliation{Department of Applied Physics, The University of Tokyo, Hongo, Bunkyo-ku, Tokyo 113-8656, Japan}
 \affiliation{Advanced Science Research Center, Japan Atomic Energy Agency, Tokai 319-1195, Japan}
\author{K. Kakurai}
 \affiliation{Neutron Science and Technology Center, Comprehensive Research Organization for Science and Society (CROSS), Tokai, Ibaraki 319-1106, Japan}
 \affiliation{RIKEN Center for Emergent Matter Science (CEMS), Saitama 351-0198, Japan}
 \affiliation{Materials Science Research Center, Japan Atomic Energy Agency, Tokai 319-1195, Japan}

\date{\today}

\begin{abstract}
We measure the mode-resolved direction of the precessional motion of the magnetic order, i.e., magnon polarization, via the chiral term of inelastic polarized neutron scattering spectra.
The magnon polarisation is important in spintronics, affecting thermodynamic properties such as the magnitude and sign of the spin Seebeck effect.
The observation of both signs of magnon polarization in Y$_3$Fe$_5$O$_{12}$ also gives direct proof of its ferrimagnetic nature.
The experiments agree very well with atomistic simulations of the scattering cross section.
\end{abstract}

\maketitle

\begin{figure*}[t]
	\includegraphics[width=0.85\textwidth]{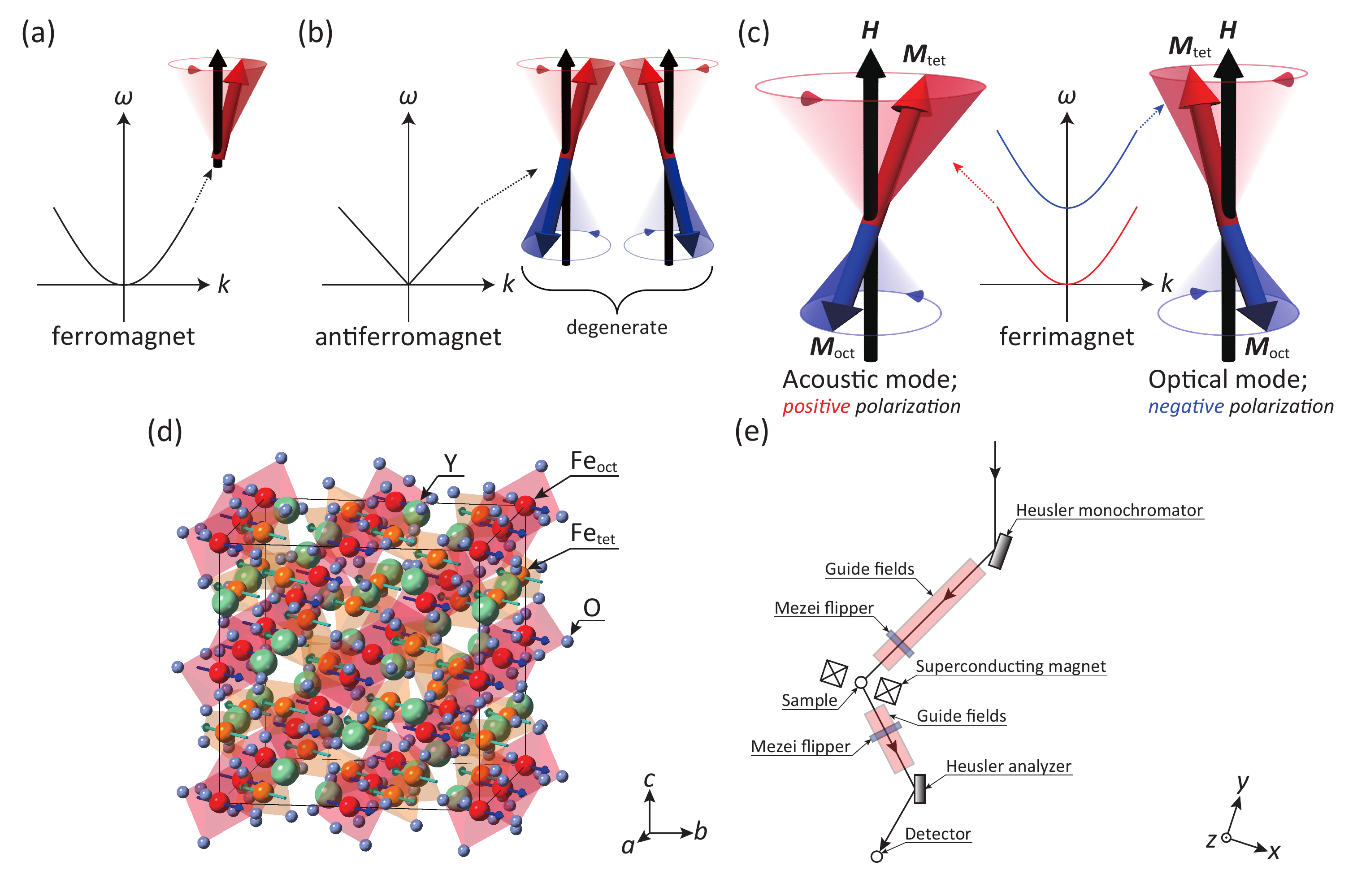}
	\caption{\label{fig1} Illustration of polarization for (a) a ferromagnet, (b) an antiferromagnet, and (c) two magnon modes in a ferrimagnet. The ``positive'' polarization acoustic mode is a coherent right-handed circular precession of the sublattice moments, whereas the ``negative'' polarization optical mode is a left-handed precession dominated by the exchange interaction between Fe$_{\rm oct}$ and Fe$_{\rm tet}$ sites. (d) Crystallographic unit cell of Y$_3$Fe$_5$O$_{12}$ with arrows marking the tetrahedral (Fe$_{\rm tet}$: $24d\ (3/8,0,1/4)$) and octahedral (Fe$_{\rm oct}$: $16a\ (0,0,0)$) sites. The magnetic moment direction is either parallel or antiparallel to the applied magnetic field direction ([110]). (e) Sketch of the IN20 instrument with bold black arrows denoting the neutron path.}
\end{figure*}

Spin waves, the elementary excitations of magnetic order in condensed matter, are quantized into ``magnons,'' bosons carrying energy, linear momentum, and spin angular momentum.
According to classical (the Landau-Lifshitz equation) and quantum mechanics, a magnetic moment precesses counter-clockwise around an applied magnetic field.
We define this motion to be ``positively'' polarized.
The collective excitations of the magnetic order in simple ferromagnets also precess only counter-clockwise; hence, all ferromagnetic magnons have a positive polarization (Fig.~\ref{fig1}(a)).
Simple collinear antiferromagnets have two magnon modes with opposite polarization (Fig.~\ref{fig1}(b)), but these are degenerate unless large magnetic fields are applied.
This impedes the possibility of observing the two polarizations.
Simple ferrimagnets have two anti-aligned sublattices and also support two magnon polarizations, but the inter-sublattice exchange field naturally separates the branches of opposite polarization into acoustic and optical modes (Fig.~\ref{fig1}(c)).
Since the energy gap between these modes can be large, spectroscopic studies have the potential to observe this polar character.
A direct experimental proof of the opposite polarization of the magnons over the exchange gap is missing, however.

The iron-based garnet Y$_3$Fe$_5$O$_{12}$ (YIG) is a ferrimagnetic insulator with a complex structure (Fig.~\ref{fig1}(d)) and is an essential material for microwave and optical technologies~\cite{Wu2013} and also for basic research in spintronics, magnonics, and quantum information~\cite{Tabuchi2015}.
One reason is that it has the highest quality magnetization dynamics among known magnets---resulting in long magnon lifetimes~\cite{Chang2014}.
The gap separating optical and acoustic modes is of the order of the thermal energy at room temperature.
A maximum of the spin Seebeck voltage in YIG near room temperature~\cite{Kikkawa2015} has been interpreted in terms of the competition between magnons of different polarization~\cite{Barker2016b}.
Even though it affects material properties, the different polarization of acoustic and optical magnon modes has never been measured.

Inelastic neutron scattering is the method of choice to measure the magnon dispersion across large areas of reciprocal space, and the magnon dispersion in YIG has previously been measured by unpolarized neutrons~\cite{Plant1977,Princep2017a,Shamoto2018}.
The unit cell contains Fe$^{3+}$ local moments with spin $S=5/2$ in tetrahedral and octahedral oxygen cages with opposite spin projection in a ratio of 3:2, giving a net magnetization.
At low temperatures, YIG behaves like a simple ferromagnet with quadratic magnon dispersion.
At higher temperatures, non-parabolicities become apparent, and optical modes start to become occupied.
In his Lectures, Feynman claimed YIG was a regular ferromagnet rather than a ferrimagnet~\cite{Feynman1963}.
This claim contradicts our general understanding~\cite{Harris1963,Cherepanov1993,Barker2016b,Rodic1999,Krichevtsova2017}, but cannot easily be refuted without a signature that can differentiate between the two.
We use polarized neutron scattering to measure the polarization of different modes in YIG.
We find negatively polarized modes over the exchange gap, as well as the positive acoustic mode, proving that YIG must be a ferrimagnet, and Feynman's assertion was incorrect.

Polarized neutron scattering has been used to separate the magnetic and nuclear contributions to scattering cross sections~\cite{Moon1969}.
It can differentiate magnon creation and annihilation processes~\cite{Moon1969}, which relates to the magnon polarization because the total spin of magnon plus neutron must be conserved in the scattering process.
The symmetry of magnetic fluctuations can also be directly investigated~\cite{Kakurai1984}.
More recently the ``chiral terms''~\cite{Maleyev1995} were used to measure chiral magnetic order~\cite{Loire2011} and excitations in paramagnetic~\cite{Roessli2002} and chiral phases~\cite{Lorenzo2007a}.
The chirality observed in these studies is a spatial variation of the {\it non-collinear} magnetic moments caused by effects such as geometrical frustration and Dzyaloshinskii--Moriya spin-orbit interactions.
Here we aim for a different property---the intrinsic polarization of the magnetic excitations in a {\it collinear} magnet.

Resolving the magnon polarization is difficult because of the low scattering intensities.
The chiral terms can only be measured when the applied field and equilibrium magnetization are aligned with the scattering wave vector ${\mathbf Q}$.
Magnetic neutron scattering can only detect the spin components perpendicular to this quantization axis, and these projections are tiny.
Besides, the signal is contaminated by imperfections in the polarizers and flippers, which are needed to select the incident and scattered neutrons (see Fig.~\ref{fig1}(e)).
We must also use high-energy (thermal) neutrons---sacrificing low momentum transfers---to reach the 60~meV energy transfers required to measure the optical modes.
A balance must be struck to maximize the scattering intensity with respect to the magnetic structure and form factors (decreasing with $Q$).
The large number of non-magnetic atoms in YIG plays in our favor because there are scattering vectors where structural (phonon) scattering is negligibly small--the nuclear structure factor is minimized--but magnetic (magnon) scattering is still present.
Measurements around the $(4,4,-4)$ reflection were anticipated to most likely lead to success.
Our first and only partially successful attempt on a different instrument (HYSPEC at Oak Ridge National Laboratory) highlights some other points of difficulty in this learning process~\cite{suppl}.

The setup of the neutron scattering instruments with the applied field parallel to ${\mathbf Q}$ (${\mathbf H}\parallel {\mathbf Q}\parallel x$), as used on the neutron triple-axis spectrometer IN20 at Institut Laue-Langevin, France, is depicted in Fig.~\ref{fig1}(e)~\cite{ILLdata}.
The scattered neutrons are recorded in four {\it channels}: $I_{x}^{++}$, $I_{x}^{--}$, $I_{x}^{+-}$, $I_{x}^{-+},$ where $I_{x}^{io}$ is the intensity of $i$ in-coming and $o$ outgoing neutrons with $+/-$ neutron polarization~\cite{Chatterji2006}.
From the four channels, the non-magnetic nuclear ($N$), magnetic ($M=M_{y}+M_{z}$), and chiral ($M_{\rm ch}$) spectra can be extracted through the combinations:
\begin{eqnarray}
N &=& \langle N_Q N_Q^{\dagger}\rangle_{\omega} = \frac{1}{2}(I_x^{++} + I_x^{--})\nonumber\\
M &=& \langle M_{Qy}M_{Qy}^{\dagger}\rangle_{\omega} + \langle M_{Qz}M_{Qz}^{\dagger}\rangle_{\omega} = \frac{1}{2}(I_x^{+-} + I_x^{-+})\nonumber\\
M_{\rm ch} &=& i(\langle M_{Qy}M_{Qz}^{\dagger}\rangle_{\omega}-\langle M_{Qz}M_{Qy}^{\dagger}\rangle_{\omega}) = \frac{1}{2}(I_x^{+-} - I_x^{-+}),\nonumber
\end{eqnarray}
where $\langle N_{Q}N_{Q}^{\dagger }\rangle _{\omega }$ and $\langle M_{Q\alpha }M_{Q\alpha }^{\dagger }\rangle _{\omega }$ $(\alpha =y,z)$ are the spatiotemporal Fourier transforms of the nuclear-nuclear and spin-spin correlation functions, respectively.
$M_{\rm ch}$ describes the chiral (or antisymmetric) correlation function within the $yz$-plane, and is proportional to the Stokes parameter~\cite{Harris1963}.
Phonon and magnon scattering have previously been separated in YIG in terms of the nuclear and magnetic spectra~\cite{Plant1977,Princep2017a}.
The chiral contribution $M_{\rm ch}$ contains the new information about the magnon polarization and forms the main result of our study.

\begin{figure}[t]
	\includegraphics[width=0.95\linewidth]{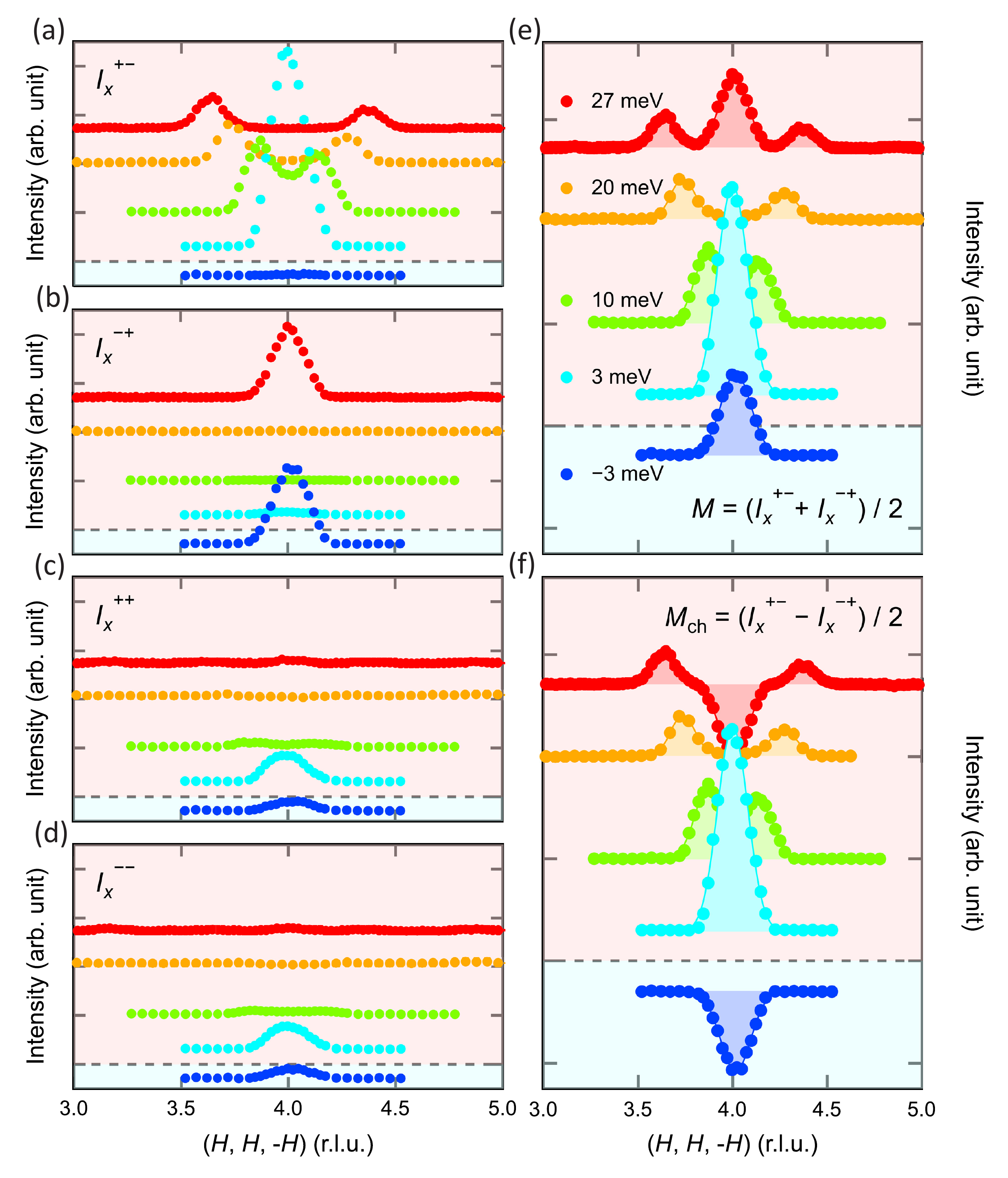}
	\caption{\label{fig2} Constant energy scans of (a) $I_x^{+-}$, (b) $I_x^{-+}$, (c) $I_x^{++}$, and (d) $I_x^{--}$ channels at representative energy transfers obtained at 293~K. Derived intensity of (e) the magnetic: $M=M_y+M_z=\frac{1}{2}(I_x^{+-}+I_x^{-+})$ and (f) chiral term: $M_{\rm ch}=\frac{1}{2}(I_x^{+-}-I_x^{-+})$. The scans run along the P$[11\bar{1}]$ direction and were taken at the fixed final wave number $k_{\rm f}=2.662$ {\AA}$^{-1}$.}
\end{figure}

Figure~\ref{fig2}(a)-(d) shows the raw data accumulated in each of the channels.
Sufficient intensity could only be obtained due to the large neutron flux available and the large spins in YIG.
The difference between $I_{x}^{+-}$ (Fig.~\ref{fig2}(a)) and $I_{x}^{-+}$ (Fig.~\ref{fig2}(b)) is immediately apparent in terms of peaks appearing in either one channel or the other.
The magnetic ($M$) and chiral ($M_{\rm ch}$) combinations are shown in Fig.~\ref{fig2}(e) and \ref{fig2}(f).
Some peaks are clearly positive and others negative, revealing the polarization of the magnon branches.

At $T=293$~K, the acoustic and optical modes are separated by a direct gap of $25$~meV at the center of the Brillouin zone, and their polarizations are opposite.
For negative energy transfer, the scattered neutrons gain energy by absorbing magnons (anti-Stokes scattering), which corresponds to an inverted polarization ($-3$~meV, dark blue in Fig.~\ref{fig2}(f)), obeying the principle of detailed balance.

\begin{figure}[t]
	\includegraphics[width=0.8\linewidth]{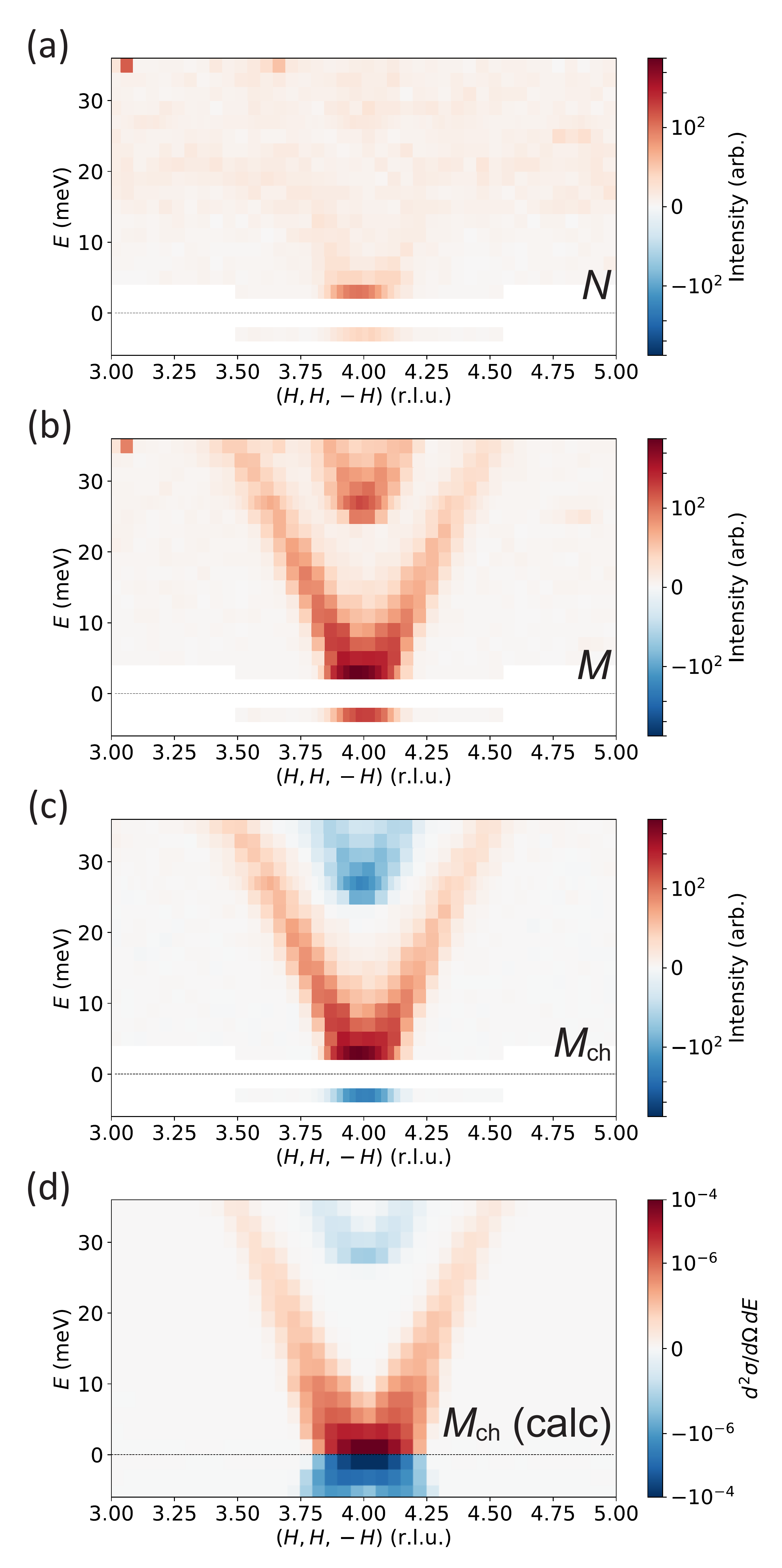}
	\caption{\label{fig3} Derived spectra below $\hbar\omega=35$~meV of (a) the nuclear: $N=\frac{1}{2}(I_x^{++}+I_x^{--})$, (b) magnetic: $M=M_y+M_z=\frac{1}{2}(I_x^{+-}+I_x^{-+})$, (c) chiral term: $M_{\rm ch}=\frac{1}{2}(I_x^{+-}-I_x^{-+})$ from mesh scans taken at 293~K with the fixed wave number $k_{\rm f}=2.662$ {\AA}$^{-1}$ around $(4,4,-4)$ in reciprocal lattice units (r.l.u.: $2\sqrt{3}\pi a^{-1}$ {\AA$^{-1}$}). Note that some scans miss the $I^{++}$ channel, which is approximated by $I^{--}$. The chiral term is compared with (d) the calculated resolution convoluted partial differential scattering cross section at 293~K~\cite{suppl}.}
\end{figure}

We summarize the results of many scans in Fig.~\ref{fig3}.
The nuclear response is very weak (Fig.~\ref{fig3}(a)) as intended: the $(4,4,-4)$ intensity is four orders of magnitude smaller than that of the strongest nuclear Bragg peak $(0,0,4)$.
The imperfections of the neutron polarization and flippers may cause the remaining weak signals, but there are almost no signatures of phonon excitations. 
The magnetic response in Fig.~\ref{fig3}(b) is equivalent to unpolarized neutron scattering, which is conventionally used to measure magnon spectra. 
The dispersion and occupation of the magnon modes are visible and agree well with previous experiments.

$M_{\rm ch}$ is plotted in Fig.~\ref{fig3}(c).
The dispersion is the same as in the magnetic response, but the sign (color) of the signal distinguishes the polarization of the modes.
The red acoustic mode has the ``positive'' polarization (counter-clockwise with respect to the field), whereas the blue optical mode is the exchange-split mode that precesses in the opposite (clockwise) direction.

We compare the measurements with the polarized neutron partial differential cross section calculated using atomistic spin dynamics with quantum statistics~\cite{Barker2016b,Barker2019,suppl}.
We convolute the calculated cross section with an approximated instrument resolution, which causes a significant broadening of the modes~\cite{suppl}.
Figure~\ref{fig3}(d) shows the simulated $M_{\rm ch}$ cross section, which shows an excellent agreement with the experiments.

\begin{figure*}[t]
	\includegraphics[width=0.95\linewidth]{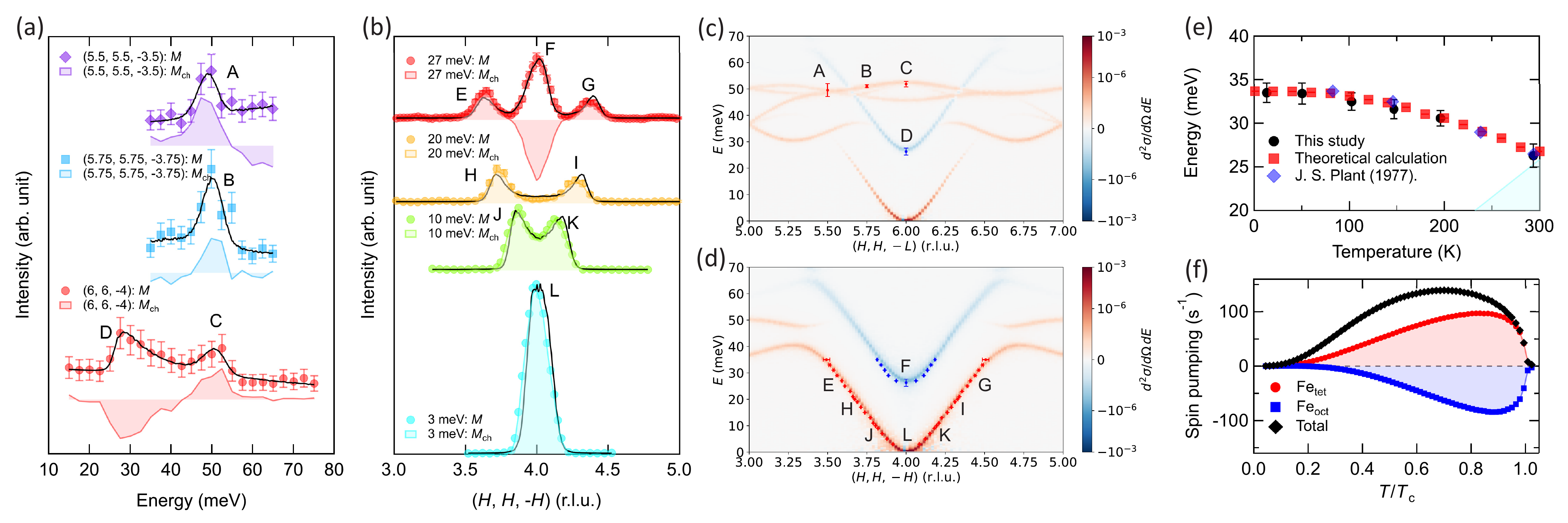}
	\caption{\label{fig4} Resolution convoluted fits for (a) constant wave number (with $k_{\rm f}=4.1$ {\AA}$^{-1}$) and (b) constant energy scans (with $k_{\rm f}=2.662$ {\AA}$^{-1}$) of the magnetic term with shaded intensity of the chiral term taken at 293~K. (c), (d) Calculated partial differential scattering cross sections overlaid with experimentally estimated peak positions. $(H,H,-L)$ in (c) and $(H,H,-H)$ in (d) span the ranges $(5,5,-3)$ to $(7,7,-5)$ and $(3,3,-3)$ to $(5,5,-5)$, respectively. (e) Temperature dependence of the estimated optical gap value compared with the calculation and the previous results~\cite{Plant1977}. The shaded area marks $E\le k_{\rm B}T$. (f) Calculated $T/T_{\rm C}$ dependence of the thermal spin pumping from Y$_3$Fe$_5$O$_{12}$ by the Fe$_{\rm tet}$, Fe$_{\rm oct}$ sites, and the total.}
\end{figure*}

We measured a large number of points on the two magnon branches and also measured an optical mode with positive polarization by moving to the $(6,6,-4)$ Brillouin zone (Fig.~\ref{fig4}(c)).
Peaks were extracted using resolution convoluted fits, which are plotted on top of the calculated scattering cross sections (Fig.~\ref{fig4}(a),\ref{fig4}(b)).
We find almost perfect agreement between experiment and theory for both the nearly flat magnon mode at 50~meV (Fig.~\ref{fig4}(c)), and the acoustic and optical modes below 35~meV (Fig.~\ref{fig4}(d)).
The magnon polarization of the localized (flat) mode is positive, in agreement with the calculations (Fig.~\ref{fig4}(a),\ref{fig4}(b)), highlighting the ability to measure polarization anywhere in reciprocal space.
The $M$ and $M_{\rm ch}$ signals in Fig.~\ref{fig4}(c),\ref{fig4}(d) are strongly correlated, giving additional confidence in the magnetic origin of $M_{\rm ch}$.

The polarization can be understood for the uniform modes--in a two sublattice macrospin model~\cite{suppl}.
One of the eigenenergies of the ferrimagnetic Hamiltonian~\cite{Schlomann1960} is zero, while the other gives the optical (exchange) gap.
The corresponding eigenoscillations are depicted in Fig~\ref{fig1}(c): the acoustic mode is a coherent right-handed circular precession in which ${\mathbf M}_{\rm tet}$ and ${\mathbf M}_{\rm oct}$ are strictly antiparallel, while the optical mode is a left-handed circular precession with finite canting angle, consistent with the observed $M_{\rm ch}$.
The gap between optical and acoustic modes in ferrimagnets is caused by the exchange field between the sublattices~\cite{Gurevich1975b}.
The optical gap closes with increasing temperature since thermal spin fluctuations reduce the sublattice magnetizations.
We measured this gap down to 10~K.
Our results agree well with previous measurements~\cite{Plant1977} and calculations~\cite{Barker2019} (Fig.~\ref{fig4}(e)).

The optical gap is important for the thermodynamic and transport properties of YIG around and above room temperature, such as the spin Seebeck effect.
Magnon modes are thermally occupied below $E=k_{\rm B}T$ (shaded area in Fig.~\ref{fig4}(e)).
At low temperatures, only the acoustic mode is occupied, but at room temperature, the optical mode with the opposite polarization plays a significant role. 
The thermal spin motive force or ``spin pumping'' that governs the spin Seebeck signal is proportional to the integrated energy times the chiral correlation function~\cite{Xiao2010} to which the acoustic and optical modes contribute with opposite sign.
This is illustrated by Fig.~\ref{fig4}(f), in which the total spin pumping signal is clearly not the sum of that from Fe$_{\rm tet}$ and Fe$_{\rm oct}$ moments, but reflects the increasing importance of the negative polarization of optical magnons on heating.
Our model is validated by experiments~\cite{Kikkawa2015}, in which the spin Seebeck voltage as a function of increasing temperature drops much faster than the magnetization.
The dc spin Seebeck effect generated by thick magnetic layers is believed to be dominated by a different spin correlation function, i.e., the Kubo formula.
Whereas a theoretical treatment is not available yet, the optical modes may be expected to play a similarly important role.
The optical modes might also explain the observation of a reduced magnon conductivity~\cite{Wimmer2018}.

To summarize, we measured the polarization of magnons in a collinear ferrimagnet and found quantitative agreement with theory.
In complex materials such as YIG, the polarization information helps to identify different modes, which is an issue as many of the 40 predicted modes cannot be observed.
We anticipate that valuable information can be gained from similar measurements on other ferrimagnets.
For example, Gd$_3$Fe$_5$O$_{12}$ shows a sign change in the spin Seebeck voltage~\cite{Geprags2016} in which modes with different polarization are thought to exist much closer together, although the large neutron absorption cross section of gadolinium will make measurements difficult.
A magnon polarization analysis of rare-earth iron garnets could also help to understand the observed magnon spin currents~\cite{Cramer2017}.
In a magnetically soft material such as YIG, the magnon polarization is nearly circular; however, YIG is very amenable to doping, and magnetic anisotropies can be introduced.
Strong anisotropies may couple magnons with opposite polarization, thereby causing ellipticity and anticrossings between optical and acoustic modes.
This ``magnon squeezing''~\cite{Kamra2016} may be essential for applications of magnets in quantum information and can be measured by this technique.
Observation of the magnon polarizations has thus demonstrated the importance of neutron scattering for the next generation of information technology with magnetic materials.

\begin{acknowledgments}
We thank M.~B\"ohm for his assistance during the experiments, and S.~Maekawa, M.~Mori, and S.~Shamoto for valuable discussions.
This work was supported by the JSPS (Nos.~16H04007, 17H05473, 19H04683, 17H06137, 16H02125, 19H00645, 25247056, 25220910, 26103006, 19K21031), ERATO ``Spin Quantum Rectification Project'' (No.~JPMJER1402) from JST, the Graduate Program in Spintronics at Tohoku University and the Royal Society through a University Research Fellowship.
Calculations were performed on ARC3, part of the High Performance Computing facilities at the University of Leeds, UK.
J.M.T. was supported at Brookhaven by the Office of Basic Energy Sciences, U.S. Department of Energy, under Contract No.~DE-SC0012704.
Work at ORNL was supported by the US-Japan Cooperative Program on Neutron Scattering.
\end{acknowledgments}

%\appendix

%\section{Appendixes}

%\nocite{*}


\begin{thebibliography}{99}
	\bibitem{Wu2013} M. Wu, A. Hoffmann, Solid State Phys. {\bf 64}, 1 (2013).
	\bibitem{Tabuchi2015} Y. Tabuchi, S. Ishino, A. Noguchi, T. Ishikawa, R. Yamazaki, K. Usami, Y. Nakamura, Science {\bf 349}, 405 (2015).
	\bibitem{Chang2014} H. Chang, P. Li, W. Zhang, T. Liu, A. Hoffmann, L. Deng, M. Wu, IEEE Mag. Lett. {\bf 5}, 6700104 (2014).
	\bibitem{Kikkawa2015} T. Kikkawa, K. Uchida, S. Daimon, Z. Qiu, Y. Shiomi, E. Saitoh, Phys. Rev. B {\bf 92}, 064413 (2015).
	\bibitem{Barker2016b} J. Barker, G.E.W. Bauer, Phys. Rev. Lett. {\bf 117}, 217201 (2016).
	\bibitem{Plant1977} J.S. Plant, J. Phys. C: Solid State Phys. {\bf 10}, 4805 (1977).
	\bibitem{Princep2017a} A.J. Princep, R.A. Ewings, S. Ward, S. T\'oth, C. Dubs, D. Prabhakaran, A.T. Boothroyd, npj Quantum Mater. {\bf 2}, 63 (2017).
	\bibitem{Shamoto2018} S. Shamoto, T.U. Ito, H. Onishi, H. Yamauchi, Y. Inamura, M. Matsuura, M. Akatsu, K. Kodama, A. Nakao, T. Moyoshi, K. Munakata, T. Ohhara, M. Nakamura, S. Ohira-Kawamura, Y. Nemoto, K. Shibata, Phys. Rev. B {\bf 97}, 054429 (2018).
	\bibitem{Feynman1963} R.P. Feynman, R.B. Leighton, M. Sands, The Feynman Lectures on Physics. (Addison-Wesley, Reading, Massachusetts, 1963), Vol. II, Chapter 37.
	\bibitem{Harris1963} A.B. Harris, Phys. Rev. {\bf 132}, 2398 (1963).
	\bibitem{Cherepanov1993} V. Cherepanov, I. Kolokolov, V. L\'vov, Phys. Rep. {\bf 229}, 81 (1993).
	\bibitem{Rodic1999} D. Rodic, M. Mitric, R. Tellgren, H. Rundlof, A. Kremenovic, J. Mag. Mag. Mat. {\bf 191}, 137 (1999).
	\bibitem{Krichevtsova2017} B.B. Krichevtsova, S.V. Gasteva, S.M. Suturina, V.V. Fedorova, A.M. Korovina, V.E. Bursiana, A.G. Banshchikova, M.P. Volkova, M. Tabuchi, N.S. Sokolova, Sci. Tech. Adv. Mat. {\bf 18}, 351 (2017).
	\bibitem{Moon1969} R.M. Moon, T. Riste, W.C. Koehler, Phys. Rev. {\bf 181}, 920 (1969).
	\bibitem{Kakurai1984} K. Kakurai, R. Pynn, B. Dorner, M. Steiner, J. Phys. C: Solid State Phys. {\bf 17}, L123 (1984).
	\bibitem{Maleyev1995} S.V. Maleyev, Phys. Rev. Lett. {\bf 75}, 4682 (1995).
	\bibitem{Loire2011} M. Loire, V. Simonet, S. Petit, K. Marty, P. Bordet, P. Lejay, J. Ollivier, M. Enderle, P. Steffens, E. Ressouche, A. Zorko, R. Ballou, Phys. Rev. Lett. {\bf 106}, 207201 (2011).
	\bibitem{Roessli2002} B. Roessli, P. B\"oni, W.E. Fischer, Y. Endoh, Phys. Rev. Lett. {\bf 88}, 237204 (2002).
	\bibitem{Lorenzo2007a} J.E. Lorenzo, C. Boullier, L.P. Regnault, U. Ammerahl, A. Revcolevschi, Phys. Rev. B {\bf 75}, 054418 (2007).
	\bibitem{suppl} See Supplemental Material at http://link.aps.org/supplemental/10.1103/PhysRevLett.?.? for additional details on experimental methods, the HYSPEC experiment, and the resolution convolution of calculated data.
	\bibitem{ILLdata} Data from project 4-01-1559 (doi:10.5291/ILL-DATA.4-01-1559)
	\bibitem{Chatterji2006} T. Chatterji, Neutron Scattering from Magnetic Materials. (Elsevier, Amsterdam, 2006).
	\bibitem{Barker2019} J. Barker, G.E.W. Bauer, Phys. Rev. B {\bf 100}, 140401(R) (2019).
	\bibitem{Schlomann1960} E. Schl\"omann, in Solid State Physics in Electronics and Telecommunication, Academic Press, New York, London, Vol. 3, p.322 (1960).
	\bibitem{Gurevich1975b} A.G. Gurevich, A.N. Anisimov, Sov. Phys. JETP {\bf 41}, 336 (1975).
	\bibitem{Xiao2010} J. Xiao, G.E.W. Bauer, K. Uchida, E. Saitoh, S. Maekawa, Phys. Rev. B {\bf 81}, 214418 (2010).
	\bibitem{Wimmer2018} T. Wimmer, M. Althammer, L. Liensberger, N. Vlietstra, S. Gepr{\"a}gs, M. Weiler, R. Gross, H. Huebl, arXiv:1812.01334.
	\bibitem{Geprags2016} S. Gepr\"ags, A. Kehlberger, F. Della Coletta, Z. Qiu, E.J. Guo, T. Schulz, C. Mix, S. Meyer, A. Kamra, M. Althammer, H. Huebl, G. Jakob, Y. Ohnuma, H. Adachi, J. Barker, S. Maekawa, G.E.W. Bauer, E. Saitoh, R. Gross, S.T.B. Goennenwein, M. Kl\"aui, Nat. Commun. {\bf 7}, 10452 (2016).
	\bibitem{Cramer2017} J. Cramer, E. Guo, S. Gepr{\"a}gs, A. Kehlberger, Y.P. Ivanov, K. Ganzhorn, F. Della Coletta, M. Althammer, H. Huebl, R. Gross, J. Kosel, M. Kl{\"a}ui, S.T.B. Goennenwein, Nano. Lett. {\bf 17}, 3334 (2017).
	\bibitem{Kamra2016} A. Kamra, W. Belzig, Phys. Rev. Lett. {\bf 116}, 146601 (2016).
\end{thebibliography}

\end{document}